\documentclass[aps,floats]{revtex4}
\usepackage{epsfig}
\usepackage{amsmath}
\usepackage{bbm}
\usepackage{graphicx}
\usepackage{color}

\begin{document}

\title{Phase diagram of the Quantum Electrodynamics of 2D and 3D Dirac semimetals}
\author{J. Gonz\'alez}
\address{Instituto de Estructura de la Materia,
        Consejo Superior de Investigaciones Cient\'{\i}ficas, Serrano 123,
        28006 Madrid, Spain}

\date{\today}

\begin{abstract}

We study the Quantum Electrodynamics of 2D and 3D Dirac semimetals by means 
of a self-consistent resolution of the Schwinger-Dyson equations, aiming to 
obtain the respective phase diagrams in terms of the relative strength of the 
Coulomb interaction and the number $N$ of Dirac fermions. In this framework, 
2D Dirac semimetals have just a strong-coupling instability characterized by 
exciton condensation (and dynamical generation of mass) that we find at a 
critical coupling well above previous theoretical estimates, 
thus explaining the absence of that instability in 
free-standing graphene samples. On the other hand, we show that 3D Dirac 
semimetals have a richer phase diagram, with a strong-coupling instability 
leading to dynamical mass generation up to $N$ = 4 and a line of critical points 
for larger values of $N$ characterized by the vanishing of the electron 
quasiparticle weight in the low-energy limit. Such a critical behavior 
signals the transition to a strongly correlated liquid, characterized 
by noninteger scaling dimensions that imply the absence of a pole in the electron 
propagator and are the signature of non-Fermi liquid behavior with no stable 
electron quasiparticles.

\end{abstract}

\maketitle



\section{Introduction}

The discovery of graphene\cite{novo} has marked the beginning of a new chapter 
in condensed matter physics, with the appearance of new fundamental concepts 
and materials with unconventional properties. We have known about other 
genuine 2D materials like the transition metal dichalcogenides, and we
have learned about the features hidden in the band structure of 
the topological insulators\cite{top1,top2}. Lately, we have also seen
the discovery of 3D Dirac semimetals which are the higher-dimensional analog 
of graphene\cite{liu,neupane,borisenko,yu}, and we have witnessed the ongoing 
search of Weyl semimetals with a built-in breakdown of parity and 
time-reversal invariance.

In the above instances, most part of the unconventional features of the 
materials come from the peculiar geometrical and topological properties of the 
band structure. Moreover, these are electron systems that are prone to being 
placed in the strong coupling regime, with a large relative strength of the 
Coulomb interaction. Graphene should be a clear example of electron system 
with strong interaction, given the large ratio between the square of the 
electron charge and the Fermi velocity in the 2D material. However, graphene 
is not a prototype of strongly correlated system, even in the case of the 
free-standing material in vacuum, as experimental observations have shown 
no sign of electronic instability down to very low doping levels\cite{exp2}.

On the other hand, most part of theoretical 
studies\cite{khves,gus,vafek,khves2,her,jur,drut1,hands2,gama,fer,ggg,sabio,prb} 
have estimated that 2D 
Dirac semimetals should have an excitonic instability at a critical point below 
the maximum interaction strength attained in graphene suspended in vacuum. The 
absence of any signature of a gap in the electronic spectrum, even below the 
meV scale, is certainly a puzzling evidence regarding the behavior of the 2D 
material. Given that the theoretical analyses have been mainly based on a 
ladder approximation to the electron self-energy corrections, the 
discrepancy between theory and experiment calls into question the use of such 
approximate methods in electron systems that are placed in the strong-coupling 
regime\cite{barnes}.

In this paper, we apply a nonperturbative approach to the investigation of 
the effects of the Coulomb interaction in both 2D and 3D semimetals, with the
aim of mapping more confidently the different phases that may appear in those 
electron systems. More precisely, we carry out the self-consistent resolution 
of the Schwinger-Dyson equations for the Quantum Electrodynamics (QED) 
of 2D and 3D Dirac semimetals, in which the scalar part of the electromagnetic 
potential is used to mediate in each case the long-range $e$-$e$ 
interaction. This approach has to be implemented in general with some kind of 
truncation to guarantee its practical feasibility. In this regard, we have 
relied on a formulation of the equations that amounts to including all kinds 
of diagrammatic contributions except those containing vertex corrections.
Nevertheless, the solutions obtained in this way account for the 
renormalization of all the quasiparticle parameters, giving rise to frequency 
and momentum-dependent forms of the quasiparticle weight, the Fermi velocity 
and the dynamical Dirac fermion mass.

In this framework, we will see that
2D Dirac semimetals have just a strong-coupling instability characterized by 
exciton condensation (and dynamical generation of mass) that we find at a 
critical coupling well above the estimates based on a ladder approximation, 
thus explaining the absence of that instability in 
free-standing graphene samples. On the other hand, we will show that 3D Dirac 
semimetals have a richer phase diagram, with a strong-coupling instability 
leading to dynamical mass generation up to $N$ = 4 and a line of critical 
points for larger values of $N$ characterized by the vanishing of the electron 
quasiparticle weight in the low-energy limit\cite{rc}. We will see that such a 
critical behavior marks the transition to a strongly correlated 
liquid, characterized by noninteger scaling dimensions that imply the absence 
of a pole in the electron propagator and are the signature of non-Fermi liquid 
behavior with no stable electron quasiparticles\cite{barnes2}.

\section{Quantum Electrodynamics of Dirac semimetals}

We focus on the QED of Dirac semimetals, for which the Fermi velocity $v_F$ 
is much smaller than the speed of light. The dynamics of these systems can 
be then described by the interaction of a number $N$ of four-component Dirac 
spinor fields $\psi_i ({\bf r})$ representing the electron quasiparticles and 
the scalar part $\phi ({\bf r})$ of the electromagnetic potential. 
In principle, each spinor can represent the electronic states around a 
different Dirac point in momentum space, but we will not make more explicit 
such a discrimination as it does not play any role in the subsequent analysis. 
The hamiltonian can be written in general as
\begin{equation}
H = i v_F \int d^D r \; \psi^{\dagger}_i ({\bf r})  \gamma_0
  \mbox{\boldmath $\gamma $}  \cdot  \mbox{\boldmath $\nabla $} \psi_i ({\bf r})   
   + e \int d^D r  \;  \psi^{\dagger}_i ({\bf r})  \psi_i ({\bf r})  
                \;  \phi ({\bf r})
\label{ham}
\end{equation}
where $\{ \gamma_\alpha \}$ is a set of Dirac matrices satisfying 
$\{\gamma_\alpha , \gamma_\beta \} = 2\eta_{\alpha \beta}$ (with the Minkowski 
metric in $D+1$ dimensions $\eta = {\rm diag}(-1,1,\ldots 1)$).

The expression (\ref{ham}) holds equally well for dimension $D = 2$ and $3$,
but the propagator of the $\phi $ field is very different in the two 
cases. The scalar field has to mediate the $e$-$e$ interaction with long-range 
Coulomb potential $V({\bf r}) = 1/|{\bf r}|$, irrespective of the spatial 
dimension. At $D = 2$, this leads to a bare propagator $D_0 ({\bf q},\omega )$
for the $\phi $ field
\begin{equation}
\left. D_0 ({\bf q})\right|_{D = 2} = \frac{1}{2 |{\bf q}|}
\end{equation}
while in 3D space the bare propagator is instead
\begin{equation}
\left. D_0 ({\bf q})\right|_{D = 3} = \frac{1}{{\bf q}^2}
\label{lr}
\end{equation}

The effects of the interaction can be characterized through the corrections 
undergone by the scalar and the Dirac field propagators. The full Dirac 
propagator $G ({\bf k},\omega_k )$ has in general a representation of the 
form
\begin{equation}
G ({\bf k},\omega_k )^{-1} = 
(\omega_k - v_F \gamma_0  \mbox{\boldmath $\gamma $} \cdot {\bf k}  ) - 
 \Sigma ({\bf k},\omega_k )            
\end{equation}
in terms of a self-energy correction $\Sigma ({\bf k},\omega_k )$ that 
contributes to renormalize the bare quasiparticle parameters. This object
is given in turn by the equation
\begin{equation}
i \Sigma ({\bf k},\omega_k ) = - e^2  \int \frac{d^D p}{(2\pi )^D} 
    \frac{d\omega_p }{2\pi } 
  D({\bf p},\omega_p )  G({\bf k} - {\bf p}, \omega_k - \omega_p) 
     \Gamma ({\bf p},\omega_p;{\bf k},\omega_k) 
\label{si}
\end{equation}
where $D({\bf p},\omega_p )$ stands for the full propagator of the 
scalar potential and $\Gamma ({\bf q},\omega_q;{\bf k},\omega_k)$ represents
the irreducible three-point vertex. More precisely, this function is defined
by the expectation value
\begin{equation}
 i e \Gamma ({\bf q},\omega_q;{\bf k},\omega_k)  =
   \langle  \phi ({\bf q},\omega_q) 
        \psi_i ({\bf k}-{\bf q},\omega_k - \omega_q) 
            \psi^{\dagger}_i ({\bf k},\omega_k) \rangle_{\rm 1PI}
\label{gamma}
\end{equation}
where 1PI means that we must take the irreducible part of the correlator.

Furthermore, $D({\bf p},\omega_p )$ has also its own equation representing it 
in terms of the full propagators and the irreducible vertex. We can write
\begin{equation}
D ({\bf q},\omega_q )^{-1} = 
 D_0 ({\bf q})^{-1} - \Pi ({\bf q},\omega_q )
\end{equation}
with the polarization $\Pi ({\bf q},\omega_q )$ being given by
\begin{equation}
i\Pi ({\bf q},\omega_q ) = N e^2 
            \int \frac{d^D p}{(2\pi )^D} \frac{d\omega_p }{2\pi } 
{\rm Tr} \left[  G({\bf q} + {\bf p}, \omega_q + \omega_p)   
   \Gamma ({\bf q},\omega_q;{\bf p},\omega_p)  G({\bf p},\omega_p ) \right]
\label{pi}
\end{equation}
Moreover, the form of the irreducible vertex 
$\Gamma ({\bf q},\omega_q;{\bf k},\omega_k)$ is constrained by the Ward
identity arising from the reduced gauge invariance of the model, admitting also 
a representation in terms of the full propagators\cite{bloch}.

The expressions (\ref{si}) and (\ref{pi}) correspond to the Schwinger-Dyson 
equations of the model. Together with a suitable representation of the 
irreducible vertex, they may lead to valuable information about the form 
of the full fermion and interaction propagators. In general, however, one has 
to resort to some kind of truncation to achieve a self-consistent resolution
of the integral equations. In what follows, we will apply a common procedure, 
the so-called bare vertex approximation, by which 
$\Gamma ({\bf q},\omega_q;{\bf k},\omega_k)$ is set equal to the unit 
matrix in the resolution of (\ref{si}) and (\ref{pi}). We note that this 
truncation does not satisfy the mentioned Ward identity, which relates the 
irreducible vertex to the derivative with respect to the frequency of the 
fermion self-energy. In this regard, the present work focuses on the
investigation of dynamical effects in the renormalization of the 
quasiparticle parameters, which could be further improved by introducing 
a suitable ansatz for the vertex (in similar fashion as in the fully relativistic 
QED\cite{bloch}). Nevertheless, we note that the relevant features reported below 
within our approach are consistent with the results found by means of 
renormalization group methods\cite{rc}, which supports the present formulation. 
The main advantage of adopting the mentioned truncation is 
that it leads to a very convenient implementation of the self-consistent 
approach, allowing us to attain easily convergence in the recursive resolution 
of the Schwinger-Dyson equations.

\begin{figure}[h]
\begin{center}
\mbox{\epsfxsize 2.0cm \epsfbox{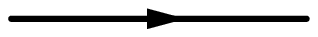}} \hspace{0.5cm} {\Large $=$}
 \hspace{0.5cm}  \mbox{\epsfxsize 2.0cm \epsfbox{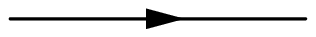}}
 \hspace{0.5cm}  {\Large $+$}  \hspace{0.5cm}
\mbox{\epsfxsize 4.8cm \epsfbox{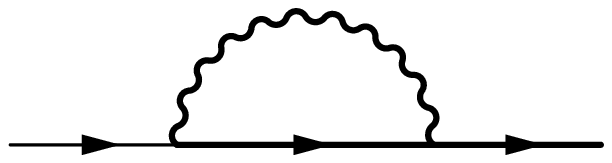}}  \\  \vspace{1cm}
\raisebox{0.2cm}{\epsfxsize 2.0cm \epsfbox{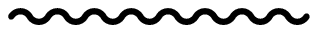}} \hspace{0.5cm} 
  \raisebox{0.16cm}{\Large $=$}
 \hspace{0.5cm}  \raisebox{0.2cm}{\epsfxsize 2.0cm \epsfbox{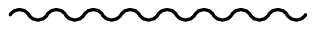}}
 \hspace{0.5cm}  
  \raisebox{0.16cm}{\Large $+$} \hspace{0.5cm}  
    \raisebox{-0.6cm}{\epsfxsize 4.8cm \epsfbox{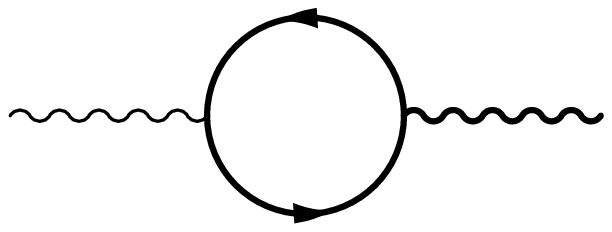}} 
\end{center}
\caption{Diagrammatic representation of the Schwinger-Dyson equations (in the bare vertex approximation). The thick(thin) straight line represents the dressed(free) Dirac fermion propagator and the thick(thin) wiggly line represents the dressed(undressed) interaction propagator.}
\label{one}
\end{figure}

Without the vertex corrections, (\ref{si}) and (\ref{pi}) lead indeed to 
closed self-consistent equations, shown diagrammatically in Fig. \ref{one}. 
This representation allows us to establish a comparison with other standard
approaches used to deal with many-body corrections. In particular, it becomes 
clear that the contributions accounted for by the diagrams in Fig. \ref{one} 
have a much more comprehensive content than other approaches dealing with the 
RPA sum of bubble diagrams for the polarization. This makes the present 
computational scheme much more reliable to describe the electron system away 
from the weak-coupling regime, incorporating effects like the renormalization 
of the Fermi velocity and the quasiparticle weight which are essential to 
capture the different critical points of the Dirac semimetals.

\section{Self-consistent resolution of Schwinger-Dyson equations}

The integral equations represented in Fig. \ref{one} can be solved numerically 
by applying a recursive procedure, after rotating first all the frequencies in 
the complex plane, $\overline{\omega} = -i \omega$, to make the passage to a 
Euclidean space in the variables $(\overline{\omega }, {\bf k})$. In practice, 
the integrals can be done numerically by discretizing the frequency and 
momentum variables. By choosing a set of frequencies 
$\overline{\omega} = \pi (2n+1)T$
with $n = 0, \pm 1, \pm 2, \ldots $, we can interpret such a discretization as
the result of placing the theory at finite temperature $T$. On the other hand,
computing with a grid in momentum space is equivalent to describing a system 
with finite spatial size. In this case, we can check the finite-size scaling 
of the results in order to extrapolate the behavior over large distances.

For the self-consistent resolution, it becomes convenient to represent the 
fermion propagator in terms of renormalization factors 
$z_\psi({\bf k}, i\overline{\omega})$ for the electron wave-function and
$z_v({\bf k}, i\overline{\omega})$ for the Fermi velocity, adding moreover 
another factor $z_m({\bf k}, i\overline{\omega})$ to allow for the dynamical 
generation of a mass for the Dirac fermions. Thus we write the full Dirac 
propagator in the form
\begin{equation}
G({\bf k}, i\overline{\omega})  =  
  \left(  z_\psi({\bf k}, i\overline{\omega}) i  \overline{\omega}  
  - z_v({\bf k}, i\overline{\omega}) v_F \gamma_0 \mbox{\boldmath $\gamma $} \cdot {\bf k}  
   -  z_m({\bf k}, i\overline{\omega}) \gamma_0 \right)^{-1}
\label{anstz}
\end{equation}
In this way, the resolution consists in finding the functions
$z_\psi({\bf k}, i\overline{\omega}), z_v({\bf k}, i\overline{\omega})$ and
$z_m({\bf k}, i\overline{\omega})$ that attain the self-consistency
in the Schwinger-Dyson equations. 

One more important detail is that the polarization $\Pi ({\bf q},\omega_q )$
may develop spurious divergences when computing the momentum integrals with 
a simple cutoff $\Lambda_k $. In general, only a gauge-invariant regularization
scheme can produce results without non-physical power-law dependences on the 
cutoff\cite{np2}. These are anyhow additive contributions to the polarization, 
which makes possible to get rid of them by a suitable subtraction procedure. 
Thus, computing with the momentum cutoff, the polarization at $D = 2$ shows a 
contribution proportional to $\Lambda_k $, while the corresponding function
at $D = 3$ has terms growing as large as $\Lambda_k^2 $. In our 
self-consistent resolution, we have carried out the frequency integrals first 
with a cutoff $\Lambda \gg v_F \Lambda_k$, implementing afterwards the
subtraction procedure to remove the power-law dependences on $\Lambda_k$ from
the polarization. In this way, we have ended up with expressions of
$\Pi ({\bf q},\omega_q )$ that are functionals of the renormalization 
factors, displaying leading behaviors at small ${\bf q}$ proportional to 
$|{\bf q}|$ and ${\bf q}^2$, respectively, for $D = 2$ and $D = 3$.

\subsection{2D Dirac semimetals}

Solving the Schwinger-Dyson equations at $D = 2$, we find in general that the 
function $z_\psi({\bf k}, i\overline{\omega})$ giving the quasiparticle weight 
remains bounded, while $z_v({\bf k}, i\overline{\omega})$ diverges in the limit 
of small momentum ${\bf k}  \rightarrow 0$. As long as the effective Fermi
velocity depends on the momentum scale, it is convenient to define the bare 
value $v_B = z_v(\Lambda_k, 0) v_F$, which can be taken as a good measure of 
the Fermi velocity at the microscopic scale (it is always verified that 
$z_\psi(\Lambda_k,0) \approx 1$). We can then define the 
unrenormalized coupling giving the bare interaction strength as
\begin{equation}
\alpha = e^2/4\pi v_B
\end{equation}
which can take different values depending on the particular Fermi velocities of 
the 2D Dirac semimetals. 

\begin{figure}[h]
\begin{center}
\includegraphics[width=3.9cm]{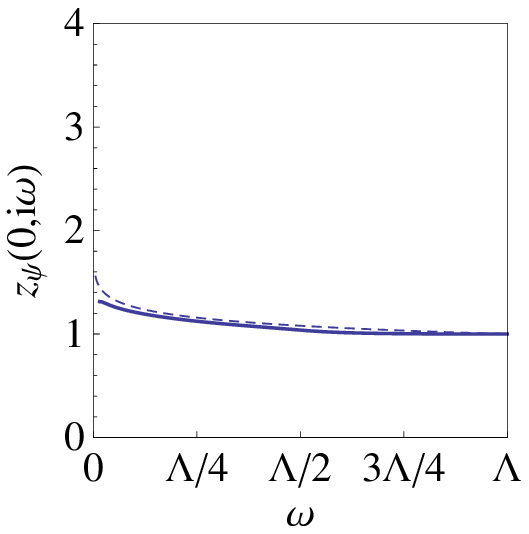}
\hspace{0.2cm}
\includegraphics[width=3.9cm]{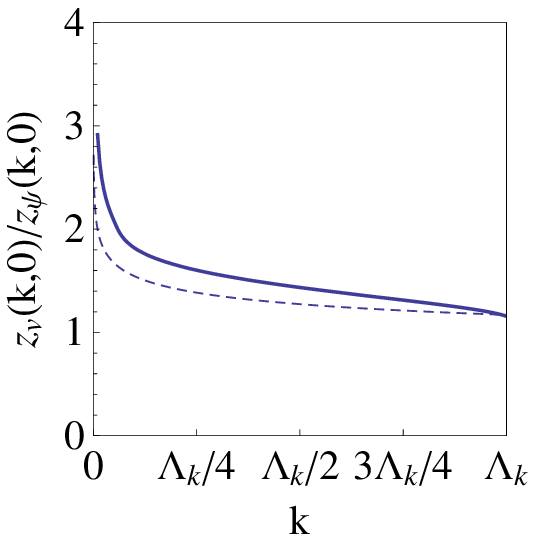}\\
 \hspace{0.36cm}  (a) \hspace{3.6cm} (b)
\end{center}
\caption{Plot of the factors $z_\psi (0,i\omega)$ and 
$z_v ({\bf k},0)/z_\psi ({\bf k},0)$ (full lines in (a) and (b)) for a 2D Dirac 
semimetal with $N = 2$, bare coupling $\alpha = 2.2$, and 
$2\pi T \approx 0.01$ eV. The dashed lines represent the dependence on the 
energy scale $\varepsilon $ of the inverse of the quasiparticle weight 
$z (\varepsilon )$ (in (a)) and the renormalized Fermi velocity 
$v (\varepsilon )$ (in (b)) obtained with the renormalization group approach
for the same bare coupling in the large-$N$ approximation.}
\label{two}
\end{figure}

As an illustration of the general behavior, Fig. \ref{two} represents the 
solution obtained for $z_\psi({\bf k}, i\overline{\omega})$ and 
$z_v({\bf k},i \overline{\omega})$ for $N = 2$ and $\alpha = 2.2$, that 
is, for parameters that should be appropriate to describe graphene 
samples suspended in vacuum. The resolution has been carried out taking a 
discretization of the frequency variables such that 
$2\pi T \approx 0.01$ eV. In this case, the self-consistency in the 
equations is only attained when 
$z_m({\bf k}, i\overline{\omega})$ is set identically equal to zero. The 
behavior found for $z_\psi({\bf k}, i\overline{\omega})$ and 
$z_v({\bf k}, i\overline{\omega})$ is in agreement with the general trend 
obtained from renormalization group methods, which found the 
divergence of the Fermi velocity in the low-energy limit as a most relevant 
feature\cite{np2,prbr}. 

For comparison, we have also represented in Fig. \ref{two} the dependence on 
the energy scale $\varepsilon $ of the inverse of the quasiparticle weight 
$z (\varepsilon )$ and the renormalized Fermi velocity 
$v (\varepsilon )$ obtained from the renormalization group approach in the 
large-$N$ approximation\cite{prbr}. It can be observed anyhow that the plot of 
$z_v ({\bf k},0)/z_\psi ({\bf k},0)$ (full line in Fig. \ref{two}(b)) follows 
the experimental results of Ref. \onlinecite{exp2} (Fig. 2(c) in that paper)
much more accurately than the scale dependence of the Fermi velocity obtained 
with the renormalization group method in the large-$N$ approximation (dashed
line in Fig. \ref{two}(b))\cite{footv}.

\begin{figure}[h]
\begin{center}
\includegraphics[width=3.9cm]{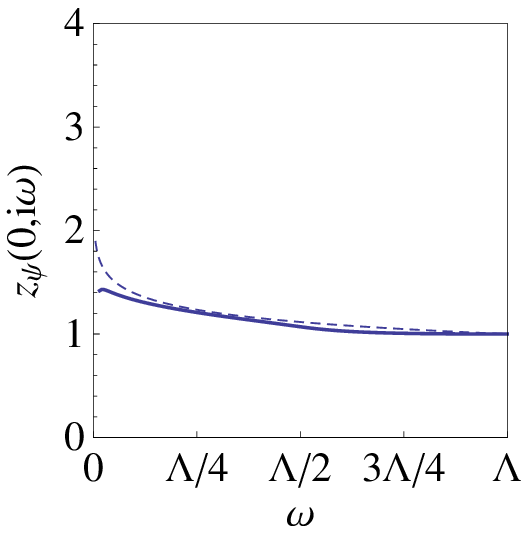}
\hspace{0.2cm}
\includegraphics[width=3.9cm]{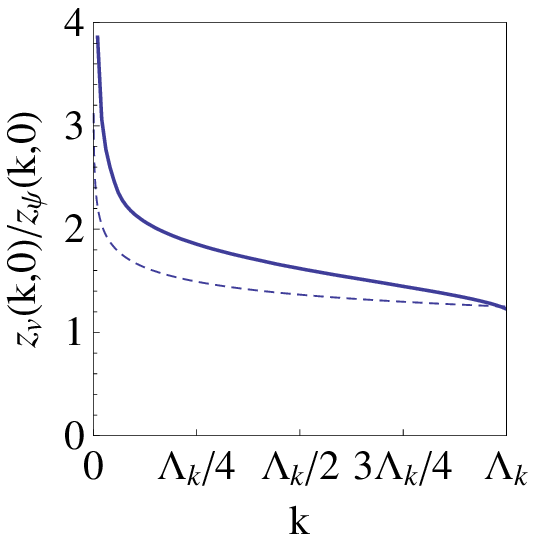}
\hspace{0.2cm}
\includegraphics[width=4.4cm]{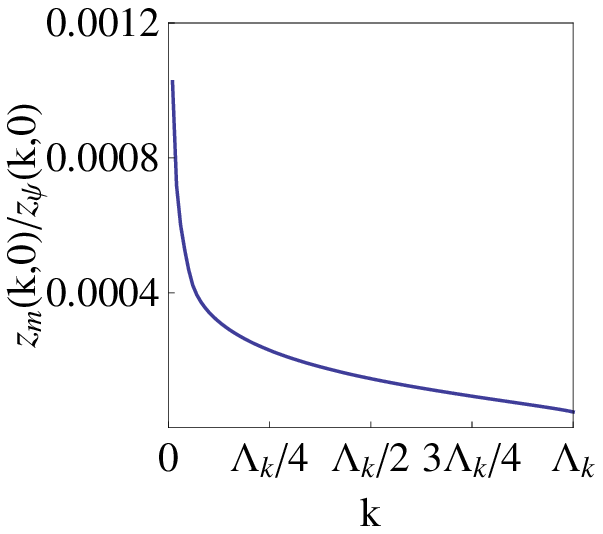}\\
  \hspace{0.36cm}  (a) \hspace{3.6cm} (b) \hspace{3.6cm} (c)
\end{center}
\caption{Plot of $z_\psi (0,i\omega)$, 
$z_v ({\bf k},0)/z_\psi ({\bf k},0)$, and 
$z_m ({\bf k},0)/z_\psi ({\bf k},0)$ measured in eV 
(full lines in (a), (b) and (c)) for a 2D Dirac semimetal 
with $N = 2$, bare coupling $\alpha = 3.37$, and 
$2\pi T \approx 0.01$ eV. The dashed lines represent the dependence on the 
energy scale $\varepsilon $ of the inverse of the quasiparticle weight 
$z (\varepsilon )$ (in (a)) and the renormalized Fermi velocity 
$v (\varepsilon )$ (in (b)) obtained with the renormalization group approach
for the same bare coupling in the large-$N$ approximation.}
\label{three}
\end{figure}

With our nonperturbative approach, moreover, we can ask whether the tendency
towards a noninteracting Fermi liquid, implied by the growth of the Fermi 
velocity, can be arrested by some instability as the bare coupling $\alpha $ 
is increased. The outcome of this search is that the other relevant feature
of the 2D Dirac semimetals is the development of a nonvanishing mass
$z_m({\bf k}, i\overline{\omega})$ at sufficiently large interaction strength,
as illustrated in Fig. \ref{three}. This corresponds to the onset of a phase 
with chiral symmetry breaking and dynamical generation of a gap for the Dirac 
fermions, as predicted by several other 
methods\cite{khves,gus,vafek,khves2,her,jur,drut1,hands2,gama,fer,ggg,sabio,prb}. 

We observe that the divergence of $z_v$ in the low-energy limit does not 
prevent the development of a nonvanishing dynamical mass $z_m$, while $z_\psi$ 
remains finite accross the transition. This latter fact implies that chiral 
symmetry breaking proceeds without an anomalous scaling of 
the Dirac fermion field, in agreement with field theory studies of that 
phenomenon\cite{nev}. In general, we also expect that the divergent growth of
the Fermi velocity is arrested at the energy scale for which the quasiparticle 
dispersion becomes gapped (which is not appreciated in the case of Fig. 
\ref{three}(b) as the infrared cutoff set in that plot by finite-size effects
is slightly above 1 meV).

\begin{figure}[h]
\begin{center}
\includegraphics[width=6.0cm]{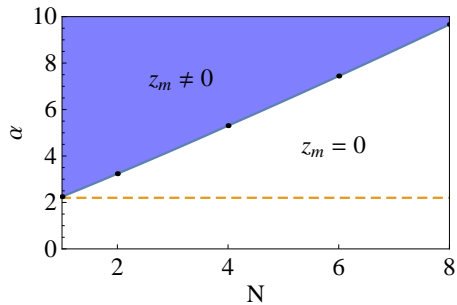}
\end{center}
\caption{Phase diagram of the QED of 2D Dirac semimetals showing the region 
with dynamically generated fermion mass $z_m \neq 0$. The critical line has been 
obtained from the self-consistent resolution of the Schwinger-Dyson
equations with $2\pi T \approx 0.01$ eV. The dashed line marks the nominal 
interaction strength $\alpha \approx 2.2$ corresponding to graphene suspended
in vacuum.}
\label{four}
\end{figure}

The present approach has the virtue of allowing an accurate
determination of the critical coupling $\alpha_c$ for the transition to the
broken symmetry phase. We have represented in Fig. \ref{four} the plot of the 
critical coupling obtained as a function of $N$, which leads to a map of the
two different phases in the QED of the 2D Dirac semimetals. In agreement with 
earlier analyses, we observe that $\alpha_c$ turns out to grow with $N$, though 
in the present resolution there seems to be no upper limit in the number of 
Dirac fermions for the development of the transition\cite{foot}. We have also 
checked that the approach to the critical coupling seems to be consistent with 
a transition of infinite order, since the dynamical mass exhibits an inflection 
point as a function of coupling constant above $\alpha_c$ which is the 
signature of that kind of transition under finite-size effects\cite{dress}.

In any case, the most relevant result regarding the phase diagram in Fig. 
\ref{four} is that the point corresponding to real graphene samples suspended
in vacuum falls in the region with no dynamical generation of mass. For the 
number $N = 2$ of Dirac fermions in graphene, the critical coupling obtained 
for chiral symmetry breaking turns out to be indeed well above most part of 
previous estimates relying on a restricted sum of many-body corrections.
The present results explain therefore that no gap has been found in the 
electronic spectrum of graphene, even in experiments looking very close to
the Dirac point\cite{exp2}. The reason for the unexpectedly large values of 
the critical coupling can be traced back to the combination of the slight 
suppression of the quasiparticle weight and the large growth of the Fermi 
velocity at low energies\cite{fv}. These two effects cooperate to reduce 
significantly the effective strength of the Coulomb interaction for the 
development of the gap, stressing the importance of a proper account of all 
the renormalization factors for the accurate determination of the transition 
to the broken symmetry phase.

\subsection{3D Dirac semimetals}

The 3D Dirac semimetals have in general a number of Dirac points that have 
attached (each of them) fermions with the two different chiralities. This is 
in particular the case of materials recently discovered like Na$_3$Bi and 
Cd$_3$As$_2$, as already clarified by their theoretical 
analysis\cite{chi1,chi2}. Such a distinctive feature of the 3D Dirac 
semimetals is relevant in the present study, since it makes possible the 
dynamical generation of mass and opening of a gap in these systems from the 
hybridization of two chiralities at the same Dirac point.

More precisely, the low-energy electronic states in both Na$_3$Bi and 
Cd$_3$As$_2$ can be naturally arranged into four-component spinors around 
each of two Dirac points, in such a way that the Dirac matrices appear in the 
chiral representation
\begin{eqnarray}
 \gamma_0 \gamma_i = 
\eta_i \left(\begin{array}{cc}
  \sigma_i  &  0  \\
 0  &  -  \sigma_i
 \end{array}\right)
\end{eqnarray}
with $\eta_1 = \eta_2 = 1$ and $\eta_3 = \pm 1$ depending on the Dirac point. 
The dynamical mass generation that we impose with the ansatz (\ref{anstz}) 
corresponds in this scheme to the mixing of the two chiralities, realized by 
the Dirac matrix
\begin{eqnarray}
\gamma_0 = 
  \left(\begin{array}{cc}
  0  &  - \mathbbm{1}  \\
 - \mathbbm{1}  &  0 
 \end{array}\right)
\label{mix}
\end{eqnarray}
A term proportional to (\ref{mix}) has been identified in Refs. \cite{chi1} 
and \cite{chi2} as one of the possible perturbations of the Dirac hamiltonian 
in Na$_3$Bi and Cd$_3$As$_2$. The physical meaning of such a term has to be
found in the breakdown of the threefold and fourfold rotational symmetry 
in each case, having an effect similar to that induced by strain in the 
crystal lattice.

\begin{figure}[h]
\begin{center}
\includegraphics[width=6.5cm]{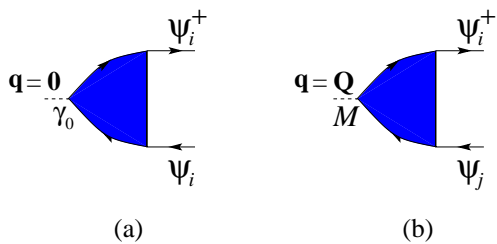}
\end{center}
\caption{Diagrammatic representation of the three-point vertices involving the
composite operators $\psi^{\dagger}_i \gamma_0 \psi_i$ and 
$\psi^{\dagger}_i M \psi_j$, with respective momentum transfer 
${\bf q} = 0$ and ${\bf q} = {\bf Q}$ connecting Dirac points.}
\label{five}
\end{figure}

The dynamical mass generation we are discussing corresponds then to the
development of an expectation value
\begin{equation}
\langle \psi^{\dagger}_i \gamma_0 \psi_i  \rangle  \neq 0
\label{dii}
\end{equation}
within each Dirac point $i$. We can imagine nevertheless the possibility of
a symmetry breaking pattern with order parameter given by 
\begin{equation}
\langle \psi^{\dagger}_i M \psi_j  \rangle  \neq 0
\label{dij}
\end{equation}
with a suitable matrix $M$ and a pair of Dirac points $i \neq j$ \cite{hut}. 
Such a condensation has the feature of involving a finite momentum ${\bf Q}$, 
needed to connect different Dirac points. This can be pictured diagrammatically, 
as the order parameters (\ref{dii}) and (\ref{dij}) can be characterized in 
terms of the respective three-point vertices in Fig. \ref{five}. The point is 
that, as implied by the renormalization group approach 
in Ref. \cite{jhep} for the 2D case, one has to hybridize the 
two chiralities at the same Dirac point so that the long-range Coulomb 
interaction may induce the strongest scaling of the three-point vertex, leading 
eventually to the condensation signaled by (\ref{dii}). The vertex in Fig. 
\ref{five}(b) is built from spinors at different Dirac points, which will not 
map in general onto each other upon a rigid shift by the momentum ${\bf Q}$ and 
may lead therefore to a weaker overlap. This means that, for a generic 3D Dirac 
semimetal with long-range Coulomb interaction, the maximum strength will be set
by the vertex in Fig. \ref{five}(a), ensuring at least the condensation given 
by (\ref{dii}) whenever symmetry breaking is to take place in the system.

\begin{figure}[h]
\begin{center}
\includegraphics[width=3.9cm]{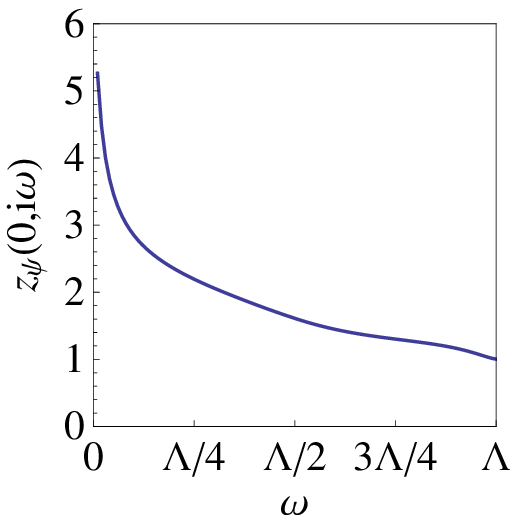}
\hspace{0.2cm}
\includegraphics[width=3.9cm]{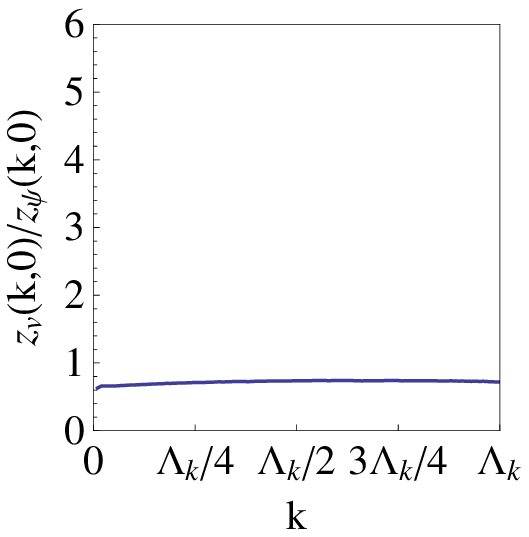}\\
 \hspace{0.36cm}  (a) \hspace{3.6cm} (b)
\end{center}
\caption{Plot of the factors $z_\psi (0,i\omega)$ and 
$z_v ({\bf k},0)/z_\psi ({\bf k},0)$ for a 3D Dirac semimetal with $N = 6$, 
bare coupling $g = 36.8$, and $2\pi T \approx 0.02$ eV.}
\label{six}
\end{figure}

Apart from the phenomenon of dynamical mass generation, the 3D Dirac semimetals
have a tendency to develop at strong coupling a drastic attenuation of the
quasiparticle weight at low energies, with a much softer renormalization of 
the Fermi velocity in comparison to their 2D Dirac analogues\cite{ros}. 
This behavior has been already found in an 
analytic study of the 3D electron systems in the large-$N$ limit, where 
it has been possible to establish rigorously the existence of a critical 
coupling at which the quasiparticle weight vanishes in the low-energy 
limit\cite{rc}. In Fig. \ref{six} we can see the effect of the critical 
behavior in the functions $z_\psi({\bf k}, i\overline{\omega})$ and 
$z_v ({\bf k}, i\overline{\omega})$, computed in the present approach for 
$N = 6$ close to the critical point and with a discretization such that 
$2\pi T \approx 0.02$ eV. For smaller values of $N$, we will see that there is 
however an interplay between that quasiparticle attenuation and the tendency 
to dynamical generation of mass. For $N \leq 4$, this latter effect becomes 
actually dominant, leading to a phase with chiral symmetry breaking that 
is the analog of the broken symmetry phase found in the 2D Dirac 
semimetals\cite{nomu}.

From the self-consistent resolution of the Schwinger-Dyson equations, we have
determined for each value of $N$ the critical coupling at which the electron
system becomes first unstable, either from the vanishing of the quasiparticle
weight or from the dynamical generation of a gap. In the case of the 3D Dirac
semimetals, one has to take special care to refer the parameters to a given
scale, since quantities like the electron charge and the Fermi velocity
are renormalized at low energies. In this respect, we have chosen to define 
the bare electron charge $e_B$ at the highest value of the momentum cutoff, 
according to the relation $e_B^2 = \Lambda_k^2 D(\Lambda_k, 0) e^2$.
Then, we can take for the microscopic parameter $e_B$ the standard 
value of the electron charge. As in the case of the 2D Dirac semimetals, we 
have also defined the bare Fermi velocity by $v_B = z_v(\Lambda_k, 0) v_F$. 
Thus, we have computed all the critical couplings referred to the relative 
interaction strength at the microscopic scale, given by
\begin{equation}
g = N e_B^2/4 \pi v_B
\label{defg}
\end{equation}

\begin{figure}[h]
\begin{center}
\includegraphics[width=6.0cm]{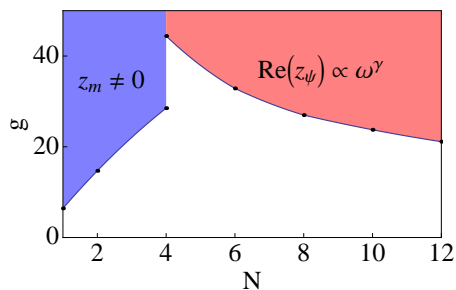}
\end{center}
\caption{Phase diagram of the QED of 3D Dirac semimetals, showing the region 
with dynamical generation of mass ($z_m \neq 0$) and the region corresponding 
to non-Fermi liquid behavior (with 
${\rm Re} (z_\psi (0, i \omega )) \propto \omega^\gamma , \; \gamma < 0$). The 
critical lines have been obtained from the self-consistent resolution of the 
Schwinger-Dyson equations with $2\pi T \approx 0.02$ eV.}
\label{seven}
\end{figure}

\begin{figure}[h]
\begin{center}
\includegraphics[width=3.9cm]{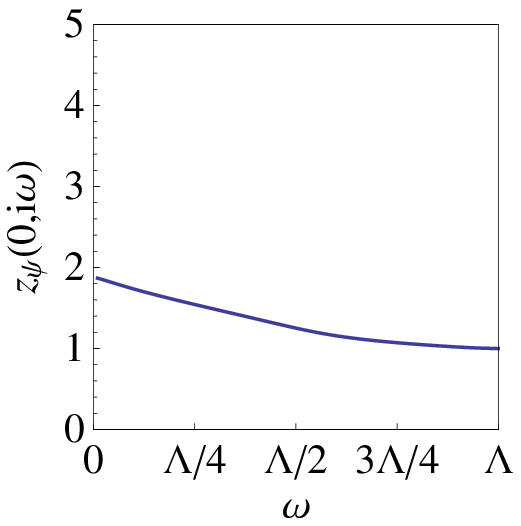}
\hspace{0.2cm}
\includegraphics[width=3.9cm]{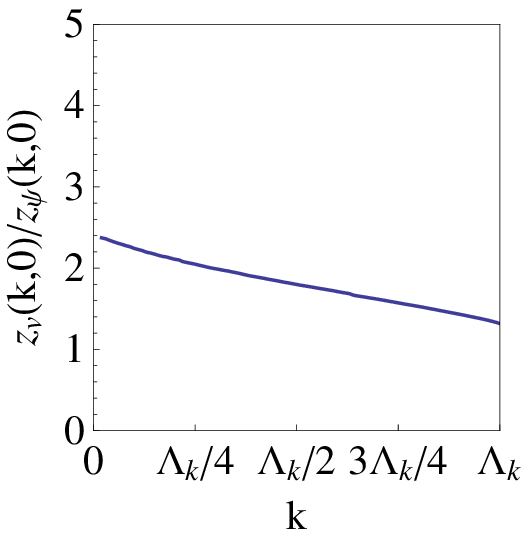}
\hspace{0.2cm}
\includegraphics[width=4.2cm]{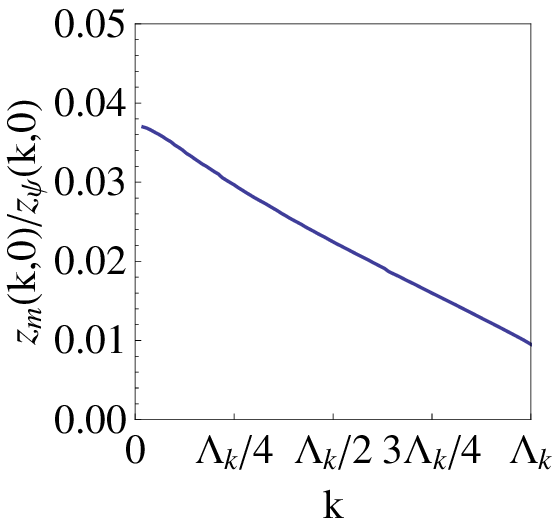}\\
 \hspace{0.36cm}  (a) \hspace{3.6cm} (b) \hspace{3.6cm} (c)
\end{center}
\caption{Plot of the factors $z_\psi (0,i\omega)$, 
$z_v ({\bf k},0)/z_\psi ({\bf k},0)$ and 
$z_m ({\bf k},0)/z_\psi ({\bf k},0)$ (measured in eV) for a 3D Dirac semimetal 
with $N = 2$, bare coupling $g = 15.1$, and $2\pi T \approx 0.02$ eV.}
\label{eight}
\end{figure}

The results we have obtained are condensed in the phase diagram shown in Fig.
\ref{seven}. We observe that there is always a phase connected to weak coupling 
for all values of $N$, characterized by a gapless spectrum of quasiparticles 
whose parameters remain regular at low energies. This phase terminates for
$N \leq 4$ in the dynamical generation of a fermion mass at sufficiently
strong coupling, which in our approach is reflected in the onset of a 
nonvanishing $z_m ({\bf k}, i\overline{\omega})$. This is shown in Fig. 
\ref{eight}, where we have represented the different renormalization factors 
for $N = 2$ at a coupling above the critical point. For this value of $N$, we 
observe that the renormalization of the quasiparticle weight is a moderate 
effect as the gap opens up in the electronic spectrum. This soft behavior 
has been also observed in the studies of dynamical mass generation in the fully 
relativistic QED, carried out by means of the self-consistent resolution of the 
corresponding Schwinger-Dyson equations\cite{bloch}.

\begin{figure}[h]
\begin{center}
\includegraphics[width=3.9cm]{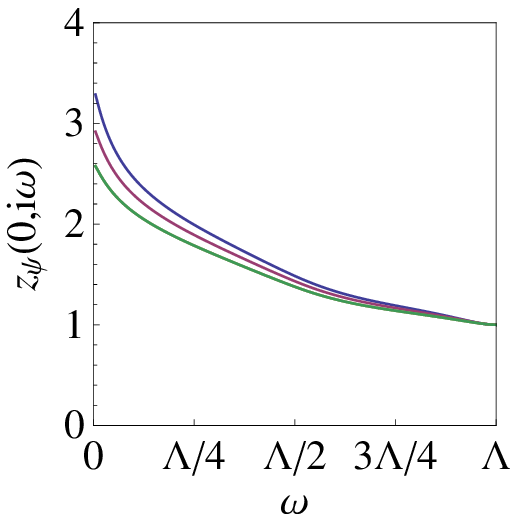}
\hspace{0.2cm}
\includegraphics[width=3.9cm]{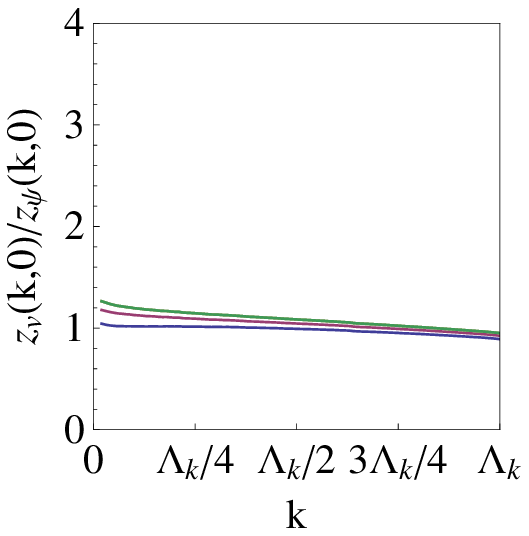}
\hspace{0.2cm}
\includegraphics[width=4.2cm]{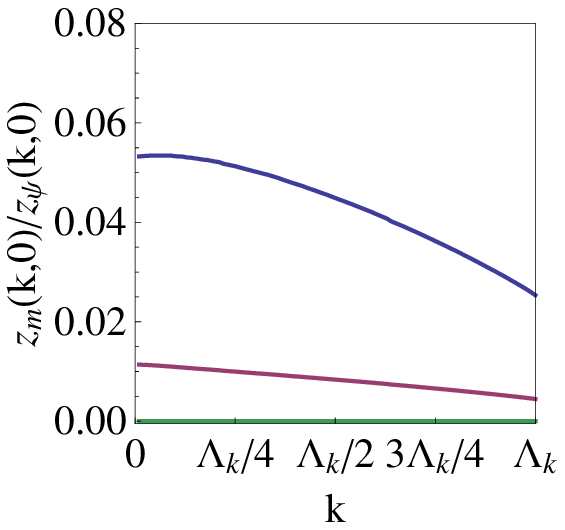}\\
 \hspace{0.36cm}  (a) \hspace{3.6cm} (b) \hspace{3.6cm} (c)
\end{center}
\caption{Plot of the factors $z_\psi (0,i\omega)$, 
$z_v ({\bf k},0)/z_\psi ({\bf k},0)$ and 
$z_m ({\bf k},0)/z_\psi ({\bf k},0)$ (in eV) for a 3D Dirac semimetal with 
$N = 4$, bare coupling $g = 30.1, 27.6$, $24.9$ 
(from top to bottom in (a) and (c), from bottom to top in (b)), and 
$2\pi T \approx 0.01$ eV.}
\label{nine}
\end{figure}

The case of $N = 4$ is however specially interesting, since there is then 
an interplay between the dynamical generation of mass and the strong 
attenuation of the quasiparticle weight. This can be observed in Fig. 
\ref{nine}, where we have plotted $z_\psi({\bf k}, i\overline{\omega})$, 
$z_v ({\bf k}, i\overline{\omega})$ and $z_m ({\bf k}, i\overline{\omega})$ for
different couplings below and above the point where the mass develops. We see
that, while the breakdown of chiral symmetry takes place before the system 
is completely destabilized by the large growth of 
$z_\psi({\bf k}, i\overline{\omega})$, this latter effect may still have a 
large impact on the observation of the quasiparticles in the electron system.

\begin{figure}[h]
\begin{center}
\includegraphics[width=3.9cm]{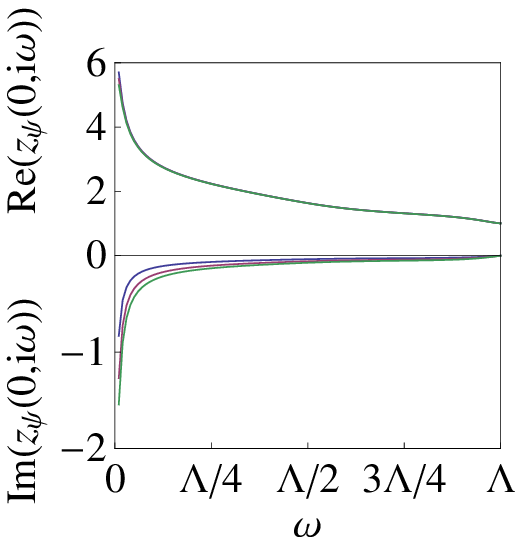}
\hspace{0.2cm}
\includegraphics[width=3.9cm]{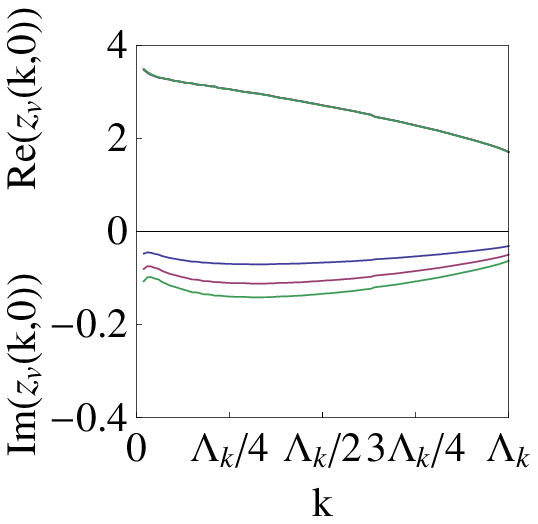}\\
 \hspace{0.36cm}  (a) \hspace{3.6cm} (b)
\end{center}
\caption{Plot of the real and imaginary parts of the factors 
$z_\psi (0,i\omega)$ and $z_v ({\bf k},0)$ for a 3D Dirac semimetal with 
$N = 6$, bare coupling $g = 37.2, 37.4$, $37.6$ (from lower 
to higher absolute value in both sides of the plot), and 
$2\pi T \approx 0.02$ eV.}
\label{ten}
\end{figure}

On the other hand, we find that for $N > 4$ there is always a critical 
coupling at which the divergence of 
$z_\psi (0,i\omega)$ takes place before the dynamical generation of mass, 
according to the trend illustrated in Fig. \ref{six}. The present approach 
allows us moreover to investigate the phase of the electron system above the 
critical point. The self-consistent resolution of the Schwinger-Dyson equations 
gives rise in that case to renormalization factors that get in general an
imaginary part, as shown in Fig. \ref{ten}. This has to be interpreted as 
the signature of a nonperturbative instability of the electron quasiparticles
since, in the conventional perturbative approach, the self-energy corrections 
obtained after Wick rotation $\omega = i \overline{\omega }$ can only account 
for the renormalization of the real part of the quasiparticle parameters. In 
our approach, the divergence of the imaginary part of 
$z_\psi (0,i\overline{\omega })$ at $\overline{\omega } = 0$ points actually 
to the development of a strongly correlated liquid, which is confirmed by the 
concomitant suppression of the quasiparticle weight in the low-energy limit, 
observed in the plots of Fig. \ref{ten}.

\begin{figure}[h]
\begin{center}
\includegraphics[width=3.9cm]{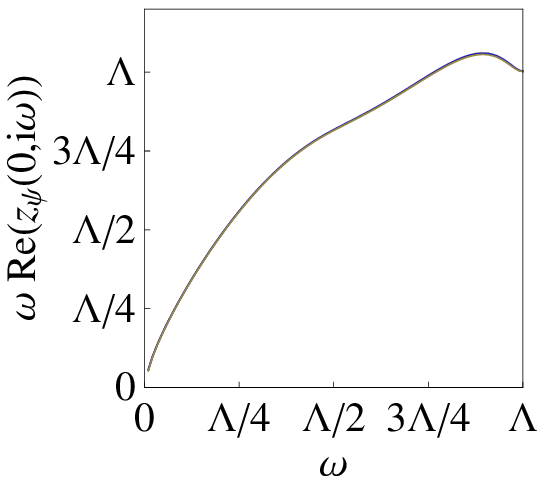}
\hspace{0.2cm}
\includegraphics[width=3.9cm]{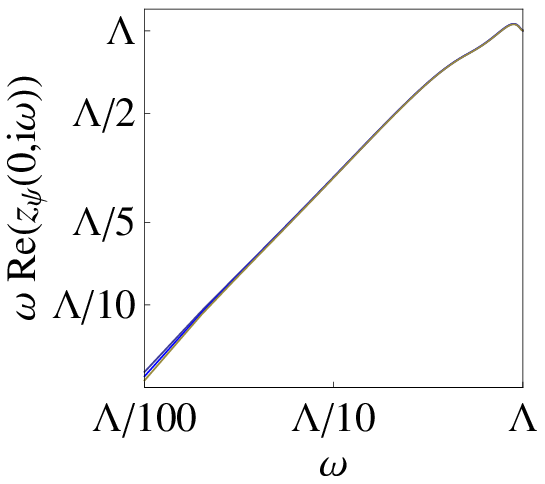}\\
 \hspace{0.36cm}  (a) \hspace{3.6cm} (b)
\end{center}
\caption{Plot of $\omega \: {\rm Re} (z_\psi (0,i\omega))$ for a 3D Dirac 
semimetal with the same parameters as in Fig. \ref{ten}, with (a) linear and 
(b) logarithmic scale (with the three different curves virtually collapsed 
onto the same line).}
\label{eleven}
\end{figure}

A detailed analysis of the renormalization factor 
$z_\psi (0,i\overline{\omega })$ in Fig. \ref{ten} reveals indeed that the 
real part of such a function follows accurately a power-law behavior as 
$\overline{\omega } \rightarrow 0$. This can be clearly seen in the 
plots of Fig. \ref{eleven}, where 
$\overline{\omega } \: {\rm Re} (z_\psi (0,i\overline{\omega }))$ is 
represented in linear and logarithmic scale. The fit to the power-law 
dependence 
\begin{equation}
\overline{\omega } \: {\rm Re} (z_\psi (0,i\overline{\omega })) \propto 
  \overline{\omega }^\mu
\end{equation}
gives an exponent $\mu \approx 0.7$, with little variation between the three
curves for the different couplings. Equivalently, this corresponds to having a 
nonvanishing anomalous dimension of the Dirac fermion field with a value of 
$\approx 0.3$. The inspection of the renormalization of $v_F$, dictated by the 
function ${\rm Re} (z_v ({\bf k},0))/{\rm Re} (z_\psi ({\bf k},0))$, leads in 
this case to a picture very similar to that in Fig. \ref{six}(b). This shows 
that $\mu $ corresponds to the exponent governing both the frequency and 
momentum scaling of the electron propagator (implying therefore a dynamical
critical exponent equal to 1).

From the physical point of view, the main consequence of the noninteger 
exponent $\mu $ is the absence of a pole in the electron propagator, which
reflects the lack of low-energy fermion excitations. This is characteristic of 
correlated systems in low dimensions, where the interactions may drive into a 
non-Fermi liquid phase with nontrivial scaling 
exponents\cite{varma,bares,nayak,hou,cast}. Our system provides 
in this respect an example of such a behavior at $D = 3$, illustrating moreover
the transition to the strongly correlated phase from the renormalized Fermi
liquid, which is the phase of 3D Dirac semimetals at sufficiently weak 
coupling.

\section{Conclusion}

We have seen that the behavior of 2D and 3D Dirac semimetals is governed by 
quite different effects in their respective strong-coupling regimes. In our
approach to the 2D semimetals, there is in general a tendency of the Fermi 
velocity of quasiparticles to grow large in the low-energy limit, in agreement
with the renormalization group studies carried out in the large-$N$ 
approximation\cite{prbr}. In the case of the 3D Dirac semimetals, we observe 
instead that the electron system is prone to develop an attenuation of the 
quasiparticle weight, with a less significant renormalization of the Fermi 
velocity. This tendency has been also identified in the large-$N$ limit of 
the 3D Dirac semimetals, which shows the existence of a critical coupling 
characterized by the divergence of the electron scaling dimension\cite{rc}.

In the 2D Dirac semimetals, the divergence of the Fermi velocity in the 
low-energy limit is the dominant feature that explains for instance the 
absence of significant correlation effects in the graphene layer. Our 
nonperturbative solution of the Schwinger-Dyson equations incorporates 
naturally the scaling of the Fermi velocity, allowing us to reach very good 
agreement with the experimental measures from graphene samples at very low 
doping levels\cite{exp2}. As a result of such a renormalization, we have 
found that the interaction strength has to be set to relatively large values, 
above those attained in the suspended graphene samples, in order to open up 
a phase with exciton condensation and dynamical generation of a gap in the 
2D Dirac semimetals.

The picture changes into a richer phase diagram for the 3D Dirac semimetals, 
as a consequence of the interplay between the attenuation of the electron 
quasiparticles that prevails at large $N$ and the tendency to dynamical mass 
generation (analogous to the chiral symmetry breaking of the fully relativistic 
QED\cite{mas,fom,fuk,mir,gus2,kon,atk,min}) that is dominant at small $N$. 
Both effects seem to coexist at the 
interface found for a number of Dirac fermions $N =4$. Most interestingly, 
our self-consistent resolution has also revealed the phase of the system above 
the large-$N$ critical point, allowing us to characterize the properties of a 
strongly correlated liquid that is reminiscent of other systems 
with suppression of electron quasiparticles making the transition from
marginal Fermi liquid\cite{varma} to non-Fermi liquid 
behavior\cite{bares,nayak,hou,cast}.

We remark that our analysis of the 3D Dirac semimetals can be extended to map
also the large-$N$ regime of 3D Weyl semimetals. These are 
a class of semimetals in which a number of Weyl points host fermions with a 
given chirality, represented in terms of two-component spinors. This means that 
a self-consistent resolution of the Schwinger-Dyson equations may be also 
carried out for these systems, writing now
the full propagator of the Weyl fermions around a given Weyl point as
\begin{equation}
G({\bf k}, i\overline{\omega})  =  
  \left(  z_\psi({\bf k}, i\overline{\omega}) i  \overline{\omega}  
  - z_v({\bf k}, i\overline{\omega}) v_F  \mbox{\boldmath $\sigma $} \cdot {\bf k}  
   \right)^{-1}
\end{equation}
We may parallel the above approach to predict the existence of a phase with 
suppression of electron quasiparticles, similar to that found for the 3D 
Dirac semimetals and covering the right part of the phase diagram in Fig. 
\ref{seven}. At small $N$ (taken now as the number of pairs of Weyl points), 
a strong-coupling phase corresponding to fermion condensation may also 
arise, with an order parameter mixing fermions at different Weyl points as in 
(\ref{dij}). The strength of this instability cannot be assessed generically, 
however, since it may be highly dependent on the particular overlapping of 
spinors from different Weyl points. Nevertheless, we may expect a 
strong-coupling symmetry broken phase for 3D Weyl semimetals at small $N$, with 
a phase boundary determined by the particular form of the spinors in 
the material hosting the Weyl points.

We finally comment on the feasibility to observe the strong-coupling phases
of the 3D semimetals, according to the values of the Fermi velocity and number 
of Dirac or Weyl points found in different materials. We recall in this regard 
that the best known examples of 3D Dirac semimetals (Na$_3$Bi and Cd$_3$As$_2$) 
have a number $N = 2$ of Dirac fermions and Fermi velocities that have been 
measured with certain accuracy\cite{liu,cd3}. The quasiparticle dispersion 
shows in both cases an anisotropy that is reflected in the values of $v_F$, 
which may be reduced by a factor of $\sim 4$ in one of the directions in momentum 
space\cite{nao}. We can make conservative estimates of the coupling defined in 
(\ref{defg}) by taking the largest Fermi velocity in each case, with the result 
that $g \sim 10$ for Na$_3$Bi and $g \sim 3$ for Cd$_3$As$_2$. These couplings 
turn out to be below the critical coupling for dynamical mass generation, 
which corresponds to the critical point at $g^* \approx 14.7$ for 
$N = 2$ in the phase diagram of Fig. \ref{seven}. This places the two mentioned 
3D Dirac semimetals in the gapless phase, showing that the transition to 
the regime with dynamical mass generation would require a Fermi velocity 
at least about $50 \%$ smaller than the largest value in Na$_3$Bi. 

On the other hand, there 
should be good prospects to observe the strong-coupling phase at large $N$ in 
Weyl semimetals. The most promising candidates for this class are the 
pyrochlore iridates and TaAs, which have 12 pairs of Weyl points. With the 
value of the Fermi velocity measured for TaAs\cite{taas}, the large value of 
$N$ already sets the effective coupling $g$ for this material well above the 
critical coupling for the non-Fermi liquid regime, which is at 
$g^*  \approx  21.1$ for 
$N = 12$. This leads us to expect that such a material should not behave as a 
regular Fermi liquid when observed at filling levels sufficiently close to the 
Weyl points. In general, we conclude that the strong-coupling phases that we 
have studied in this paper are not beyond reach, and that they may be found 
in systems with reasonably low Fermi velocities, or with a sufficiently large 
number of Dirac or Weyl points already exhibited by several materials.

\vspace{0.5cm}

We acknowledge the financial support from MICINN (Spain) through grant 
FIS2011-23713 and from MINECO (Spain) through grant FIS2014-57432-P.


\begin{thebibliography}{24}
\bibitem{footv}
In the nonrelativistic approximation, the Fermi velocity diverges without bound 
in the low-energy limit as a consequence of dealing with the static scalar 
interaction, since it has been shown in Ref. \cite{np2} that $v(\varepsilon )$
actually converges to the speed of light in the theory with the 
full (retarded) gauge interaction.
\bibitem{foot}
This feature is also present in the refined resolution of the gap equation
with dynamical screening carried out in Ref. \onlinecite{ggg}.
\end{thebibliography}
\end{document}